\documentclass[preprint,proceedings]{rmaa}

\suppressfulladdresses 

\usepackage{paralist}

\usepackage{psfrag,color}

\SetYear{2007}
\SetConfTitle{Massive Stars: Fundamental Parameters and Circumstellar Interactions}

\title{Massive Stars: From the VLT to the ELT}

\author{Chris Evans \altaffilmark{1}}

\altaffiltext{1}{UK Astronomy Technology Centre, 
Royal Observatory, Blackford Hill, Edinburgh, EH9 3HJ, UK
(cje@roe.ac.uk).}

\shortauthor{Chris Evans}

\listofauthors{C. J. Evans}
\indexauthor{Evans, C. J.}

\abstract{
New facilities and technologies have advanced our understanding of
massive stars significantly over the past 30 years.  Here I introduce
a new large survey of massive stars using VLT-FLAMES, noting the
target fields and observed binary fractions.  These data have been
used for the first empirical test of the metallicity dependence of the
intensity of stellar winds, finding good agreement with theory -- an
important result for the evolutionary models that are used to
interpret distant clusters, starbursts, and star-forming galaxies.
Looking ahead, plans for future Extremely Large Telescopes (ELTs) are
now undergoing significant development, and offer the exciting
prospect of observing spatially-resolved massive stars well beyond the
Local Group.  
}
\resumen{
Nuestro conocimiento de las estrellas masivas ha aumentado
significativamente en los \'ultimos 30 a\~nos gracias a las nuevas
instalaciones y tecnolog\'ias.  En esta contribuci\'on presento un
gran {\em survey} de estrellas masivas que se ha realizado
recientemente mediante el uso de VLT-FLAMES, mostrando los campos
observados y remarcando la fracci\'on de estrellas binarias que se ha
encontrado. Estos datos se han utilizado para la primera
comprobaci\'on empirica de la dependencia de la intensidad de los
vientos estelares con la metalicidad, encontrandose un buen acuerdo
con la teoría -- un resultado de gran importancia para los modelos de
evoluci\'on estelar, que se utilizan para la interpretai\'on de
c\'umulos lejanos, {\em starburst} y galaxias con formaci\'on estelar.
Dando un paso m\'as, comentar\'e como en la actualidad se est\'an
dedicando grandes esfuerzos al avance de los planes de actuación y
desarrollo de los Telescopios de Gran Tama\~no, que ser\'an una realidad
en un futuro pr\'oximo; este hecho nos ofrecer\'a una posibilidad
m\'as que interesante para obtener observaciones con resoluci\'on
espacial de estrellas masivas m\'as all\'a del Grupo Local.
}

\addkeyword{Galaxies: Magellanic Clouds}
\addkeyword{Instrumentation}
\addkeyword{Stars: early type}

\begin{document}
\maketitle

\section{Introduction}
\label{sec:intro}
An excellent example of the progress enabled by new observing
capabilities is afforded by our knowledge of the stellar content of
NGC\,346 -- the largest H~{\scriptsize II} region in the Small Magellanic
Cloud (SMC).

The first photometric study of NGC\,346 was published by Niemela et
al. (1986) using data from the 1-m Yale telescope at the Cerro Tololo
Inter-American Observatory (CTIO).  This paper also presented
spectroscopy of some of the brighter members, building on the
pioneering work of Walborn (1978).  In the following years, 4-m class
telescopes were used to investigate the cluster initial mass function 
(Massey et al., 1989), and to obtain high-resolution echelle
spectra for detailed atmospheric analysis (Kudritzki et al. 1989;
Walborn et al. 2000; Bouret et al.  2003).  More recently, the images
of the cluster from the {\it Hubble Space Telescope} Advanced Camera
for Surveys (e.g. Sabbi et al. 2007) offer a dramatic illustration of
the current `state-of-the-art'.

Understanding the role of environment and metallicity on the evolution
of massive stars has been a key topic for the past two
decades.  However, until now, we have lacked the facilities to obtain
large, homogenous sets of observations (in a sensible allocation of
telescope time) to provide robust empirical constraints to
evolutionary models.  The delivery of the Fibre Large Array
Multi-Element Spectrograph (FLAMES) to the Very Large Telescope (VLT)
was the catalyst for such a survey -- to address questions such as the
metallicity dependence of stellar rotational velocities and
wind mass-loss rates.

\section{The VLT-FLAMES Survey}
The FLAMES survey of massive stars has observed over 800 targets in 7
fields, centered on stellar clusters in the Galaxy and SMC/LMC, as
listed in Table~1; the inclusion of NGC\,346 was an obvious choice.
The full content of the survey has been presented by Evans et
al. (2005, 2006).

Our observations provide lower limits to the binary
fraction of the O- and early B-type targets, finding $\sim$25$\%$ in
NGC\,346 and $\sim$35$\%$ in N11.  Although the programme was not
originally concerned with binary detection, the FLAMES spectra are of
sufficient quality that the nature and properties of many of the newly
discovered systems can be determined.  Specific systems will be the
subject of future papers, and we are seeking further monitoring of our
fields to better constrain the binary fraction.

\begin{table}[t]
\caption{Summary of VLT-FLAMES fields \label{fields}}
\begin{center}
\begin{tabular}{lll}
\hline
& `Young' clusters & `Old' clusters \\
& ($<$5 Myrs)      & (10-20 Myr)    \\
\hline
Milky Way & NGC 6611 & NGC 3293 \& 4755 \\
LMC       & N11 & NGC 2004 \\
SMC       & NGC 346 & NGC 330 \\
\hline
\end{tabular}
\end{center}
\end{table}

The first science papers from the survey have now been published, 
presenting analyses of the O- and early B-type spectra (Mokiem et
al. 2006, 2007; Hunter et al. 2007).  One of our primary motivations
for the FLAMES survey was to test the theoretical prediction that wind mass-loss rates
are dependent on metallicity (Kudritzki et al. 1987; Vink et al. 2001).  
Figure~1 shows the wind momentum -- luminosity relations
(WLR) obtained from our LMC and SMC targets, compared with those from
contemporary Galactic results.  This is the first comprehensive
empirical test of the metallicity dependence.  For luminosities
greater than $\sim$10$^{5.2}$~L$_{\odot}$, the relative offsets
between different metallicity regimes are in good agreement with
theory.

\section{ELT Science Case}
With instruments such as the FOcal-Reducer low-dispersion Spectrograph
(FORS) on the VLT, detailed analysis of stars in galaxies at 1-2~Mpc
is currently possible (e.g. Urbaneja et al., 2003; Evans et al.,
2007).  But ideally we want to reach out to more distant systems (thereby
sampling different environments -- elliptical galaxies, lower
metallicities etc.), or we need spectral resolutions that are greater than
those available from FORS.

Moving into 2007, the European Southern Observatory (ESO) is starting
a full phase A design of its Extremely Large Telescope (ELT).  The
current baseline reference design is a 42-m primary, with a novel
5-mirror solution that provides correction for ground-layer
atmospheric turbulence as its minimum operating mode.  An ELT will
allow us to observe the resolved massive-star populations in a wide
range of systems beyond the Local Group, exploring chemical
abundances, stellar kinematics and so on.  An ELT will also be able to
probe distant clusters and starbursts at new levels of detail to
study their initial mass functions and kinematic structure.

\begin{figure}[t]
\includegraphics[width=\columnwidth]{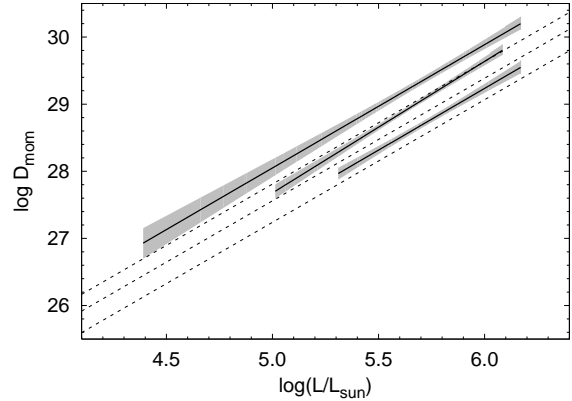} 
\caption{The observed wind momentum -- luminosity relations
(solid lines) compared with theoretical predictions (Vink et al., 2001;
dotted lines).  The top, middle and bottom lines correspond, respectively, to
Galactic, LMC and SMC results (Mokiem et al., submitted).}
\end{figure}

There will be many exciting new opportunities in the ELT era, but it is
worth noting for our community that the most significant observational
advances will almost certainly arise in the near-infrared (because of
the demands of adaptive optics).  To fully exploit an ELT, a large
effort will be required to develop further diagnostic tools in this
wavelength region -- both in terms of atomic data (e.g. Przybilla,
2005), and comparison studies in the Milky Way and Magellanic Clouds.

\end{document}